\documentclass[singlecolumn,preprint,prb,nofootinbib]{revtex4-2}
\usepackage{graphicx} 
\usepackage{amsmath}
\usepackage{amssymb}
\usepackage{hyperref}
\usepackage{pgfplots}
\usepackage{tikzviolinplots}
\usepackage{pgfplotstable}
\newcommand{\iu}{\mathrm{i}\mkern1mu}

\pgfplotsset{compat=1.18}
\usepgfplotslibrary{external}
\usepgfplotslibrary{fillbetween}
\date{December 2023}

\begin{abstract}
    The $GW$ approximation has become an important tool for predicting charged excitations of isolated molecules and condensed systems. Its popularity can be attributed to many factors, including a favorable scaling and relatively good accuracy. In practical applications, the $GW$ is often performed as a one-shot perturbation known as $G_0W_0$. Unfortunately, $G_0W_0$ suffers from a strong starting point dependence and is often not as accurate as one would need. Self-consistent $GW$ methodologies alleviate these problems but come with a marked increase in computational cost. In this manuscript, we propose the use of an estimate of the exchange-correlation derivative discontinuity to provide a remarkably good starting point for $G_0W_0$ calculations, yielding ionization potentials and electron affinities with eigenvalue self-consistent $GW$ quality at no additional cost. We assess the quality of the resulting methodology with the GW100 benchmark set and compare its advantages over other similar methods.
\end{abstract}

\begin{document}
\title{Exploiting a derivative discontinuity estimate for accurate $G_0W_0$ ionization potentials and electron affinities}
\author{Daniel Mejia-Rodriguez}
\affiliation{Physical and Chemical Sciences Division, Pacific Northwest National Laboratory, Richland WA 99352}

\maketitle

\section{Introduction}
Density functional theory (DFT)\cite{Hohenberg1964,Kohn1965} is now recognized as one of the more important tools for determining the electronic structure of molecules and extended systems. It is well-known, however, that common density functional approximations (DFAs) possess some shortcomings attributable to the limitations and constraints of such approximations \cite{Ruzsinszky2011}. For example, it has been known for decades that the Kohn-Sham eigenvalue of the highest occupied molecular orbital (HOMO) is the negative of the vertical ionization potential (VIP), $-I$ \cite{Perdew1982,Perdew1983,PerdewLevy1997}. Common DFAs, however, yield HOMO eigenvalues that deviate substantially from $-I$ due to the incorrect continuous behavior of the energy as a function of the number of electrons $N$, instead of the correct piecewise linear relation \cite{Perdew1982,Perdew1983,Ayers2008}. The eigenvalue of the lowest unoccupied molecular orbital (LUMO) also shows important deviations from the negative of the vertical electron affinity (VEA), $-A$, due to the deviation from the linearity condition \cite{Cohen2008,Cohen2012}. These undesirable behaviors have been identified as fractional charge errors which translate into either delocalization (convex deviations) or localization (concave deviations) errors \cite{Mori2008,Cohen2012,Prokopiu2022}. It has been found that DFAs based on the local density approximation (LDA), generalized gradient approximation (GGA), or meta-GGA show a rather large convex curve for the energy as a function of $N$, while Hartree-Fock (HF) yields a concave deviation, a fact that has been exploited in the development of several DFAs including optimally-tuned range-separated hybrids \cite{Baer2005,Stein2010,Kronik2012,Refaely2012,Autschbach2014,Jin2016,Mussard2017,Kronik2020}.
Unfortunately, the errors in approximating VIP and VEA through the frontier KS eigenvalues do not cancel out as most of the time VIP is underestimated and VEA is overestimated, usually resulting in the significant underestimation of the fundamental gap $E_g = I - A$ in solids (bandgap) and molecules (hardness \cite{Parr1983}). This has been known as the \emph{bandgap problem} (of DFT) and affects the accuracy of several response properties and chemical reactivity descriptors \cite{Geerlings2020} that depend on $E_g$. 

KS eigenvalues from orbitals below HOMO and above LUMO have also been used as approximate addition and removal energies \cite{Vargas2005,Gritsenko2009,Dabo2009,Dabo2010,Tsuneda2010,Korzdorfer2012,Refaely2012,Dabo2013,Ferretti2014,Borghi2014,Egger2014,Nguyen2015,Gallandi2015,Nguyen2016,Gallandi2016,Jin2016,Nguyen2018,Mei2019,Hirao2020,deGennaro2022,Colonna2022}. There is no rigorous theoretical connection between the KS eigenvalues and the quasiparticle energies, but there are practical benefits of a DFA that yields eigenvalues in reasonably close agreement with quasiparticle energies. On the one hand, these KS eigenvalues could be directly used to try to understand and predict the properties of novel materials in a computationally efficient way. On the other hand, both KS eigenvalues and orbitals can be used as input for other many-body methods known to solve, or at least alleviate, the bandgap problem. Prominent among these methods is the $GW$ approximation (GWA) to the self-energy \cite{Hedin1965,Aryasetiawan1998,Onida2002}. The GWA is usually performed perturbatively as a one-shot correction, $G_0W_0$ \cite{Hybertsen1985,Hybertsen1986}, which can be significantly affected by the quality of the starting orbitals and eigenvalues \cite{Bruneval2012,Zhang2022}. Fully self-consistent $GW$ (sc$GW$) \cite{Holm1998,Caruso2012} calculations provide one way to eliminate the starting-point dependence at a steep computational cost. Lower-scaling partial self-consistent $GW$ approaches, like quasiparticle self-consistent $GW$ (qs$GW$) \cite{Faleev2004,Schilfgaarde2006} and eigenvalue self-consistent $GW$ (ev$GW$) \cite{Shishkin2007,Stan2009}, reduce the starting-point dependence but still have much higher computational costs than $G_0W_0$. It is therefore highly desirable to find a scheme to achieve accurate $G_0W_0$@DFA with reasonable computational costs and robust accuracy.

Various efforts have been made in this direction, including the use of the \emph{renormalized singles} Green's function ($G_{RS}W_0$, $G_{RS}W_{RS}$, $G_{RSc}W_{0}$, $G_{RSc}W_{RSc}$) \cite{Jin2019:1,Li2022,Li2022:2}, the development of Koopman's compliant DFA \cite{Dabo2009,Dabo2010,Dabo2013,Ferretti2014,Borghi2014,Borghi2015,Nguyen2015,Nguyen2016,Nguyen2018,Colonna2019,deGennaro2022,Colonna2022}, and the development of optimally-tuned range-separated hybrid functionals \cite{Refaely2012,Egger2014,Gallandi2015,Gallandi2016,Bois2017,McKeon2022}. Here, we propose a fourth route using corrected eigenvalues from a family of recent generalized-gradient approximation (GGA) with nearly correct asymptotic potential (NCAP)\cite{Carmona2019,Carmona2020:1,Carmona2022:1}. This route is similar in spirit to the $\bar{\Delta}GW$ method from Vl\v{c}ek and cols. \cite{Vlcek2018}, but here we use a shift \emph{before} the $GW$ calculation as the NCAP DFAs exploit the fact that the derivative discontinuity (DD) implied in some DFAs as $N$ crosses integer values can be directly obtained from the non-zero constant needed to make the exchange-correlation (XC) potential, $v_{\text{xc}}$, vanish at infinity. We will show that NCAP eigenvalues corrected by the DD shift (NCAP-DD) provide a remarkably good starting point for $G_0W_0$ calculations, yielding quasiparticle energies on par with the more expensive partially self-consistent ev$GW$@GGA and ev$GW$@mGGA. We will begin our presentation by introducing the theory behind the development of the NCAP functionals, followed by a basic introduction to the GWA. Then, we will use the GW100 \cite{vanSetten2015} benchmark set to showcase the accuracy that can be achieved by the $G_0W_0$@NCAP-DD method.

\section{Theory \label{sec:theory}}
\subsection{Derivative discontinuity}
The exact behavior of the energy, $E$, as a function of the number of electrons is given by lines connecting the integer values of $N$ \cite{Perdew1982,Perdew1983,Ayers2008}, generally with different slopes on the electron-deficient and electron-abundant sides of $N$. This piecewise linear behavior implies that
\begin{equation}
    E_g = \left( \frac{\partial E}{\partial N} \right)_{v(\mathbf{r})}^+ - \left( \frac{\partial E}{\partial N} \right)_{v(\mathbf{r})}^- = I - A \; , \label{eq:1}
\end{equation}
for the exact functional, where the superscripts indicate that the derivative is obtained from the electron abundant ($+$) or deficient side ($-$). Equation (\ref{eq:1}), in turn, implies that the functional derivative of $E_{\text{xc}}$ with respect to the electron number density, $n(\mathbf{r})$, is also discontinuous as $N$ crosses an integer value \cite{Perdew1983,Sham1983,Kohn1986,PerdewLevy1997}, i.e.
\begin{equation}
    \Delta_{\text{xc}}(\mathbf{r}) = v_{\text{xc}}^{+}(\mathbf{r}) - v_{\text{xc}}^{-}(\mathbf{r}) \neq 0 \; .
\end{equation}
It is customary to assume that the DD $\Delta_{\text{xc}}$ is constant \cite{Sagvolden2008} and that
\begin{equation}
    \Delta_{\text{xc}} = \varepsilon_{\text{H}}^{+} - \varepsilon_{\text{L}}^{-} \; ,
\end{equation}
with $\varepsilon_{\text{H}}^{+}$ as the HOMO eigenvalue determined with $v_{\text{xc}}^{+}(\mathbf{r})$, and $\varepsilon_{\text{L}}^{-}$ as the corresponding LUMO eigenvalue determined with $v_{\text{xc}}^{-}(\mathbf{r})$. Furthermore, the ionization potential theorem states that \cite{PerdewLevy1997}
\begin{equation}
    \varepsilon_{\text{H}}^{-} = -I \;\; \text{and} \;\; \varepsilon_{\text{H}}^{+} = -A \; ,
\end{equation}
where $\varepsilon_{\text{H}}^{-}$ is the HOMO eigenvalue determined with $v_{\text{xc}}^{-}(\mathbf{r})$. Several relations can be obtained from the previous set of equations, including the confirmation that the KS gap obtained from XC DFAs that do not incorporate the DD of the XC potential will underestimate the fundamental gap
\begin{equation}
    E_g = \varepsilon_{\text{L}}^- - \varepsilon_{\text{H}}^- + \Delta_{\text{xc}} = E_g^{\text{KS}} + \Delta_{\text{xc}}
\end{equation}

Moreover, the shifts needed to pair the Kohn-Sham eigenvalues to the negative VIP and VEA
\begin{equation}
    \Delta_{\text{xc}}^- = -(\varepsilon_{\text{H}}^- + I) \;\; \text{and} \;\; \Delta_{\text{xc}}^+ = -(\varepsilon_{\text{L}}^- + A) 
\end{equation}
can be related to the full DD via
\begin{equation}
    \Delta_{\text{xc}} = \Delta_{\text{xc}}^+ - \Delta_{\text{xc}}^-
\end{equation}
The takeaway piece of this analysis is that the full DD can be decomposed into separate shifts for the occupied and unoccupied spaces, a property exploited by Carmona-Esp\'indola and cols. \cite{Carmona2020:1,Carmona2022:1}

\subsection{The NCAP exchange functional}
The NCAP exchange enhancement factor \cite{Carmona2019}
\begin{equation}
    F_x(s) = 1 + \mu \, \tanh(s) \sinh^{-1}(s) \frac{1 + \alpha \, \left((1-\zeta)s\ln(1+s) + \zeta s \right) }{1 + \beta \, \tanh(s) \sinh^{-1}(s)} \; ,
    \label{eq:ncap}
\end{equation}
given in terms of the reduced density gradient $s(\mathbf{r}) = |\nabla n(\mathbf{r}) | / 2(3\pi^2)^{1/3} n^{4/3}(\mathbf{r})$,
was purposely designed to generate an exchange potential that goes asymptotically to a constant plus a term that decays as $-c/r$. The values of the constants appearing in Equation (\ref{eq:ncap}) were obtained by satisfaction of different constraints or norms and are given in Table \ref{tab:1}

\begin{table}[h]
\center
\caption{Constants defining the NCAP and NCAPR exchange enhancement factors.}
\label{tab:1}
\begin{ruledtabular}
\begin{tabular}{c  c c }
                 & NCAP & NCAPR \\\hline
   $\alpha$      & 0.345117 & 0.343152 \\
   $\beta$       & 0.018086 & 0.017983 \\
   $\mu$         & 0.219515 & 0.219515 \\
   $\zeta$       & 0.304121 & 0.500000
\end{tabular}
\end{ruledtabular}
\end{table}

It can be shown that for a density whose asymptotic behavior is given by
\begin{equation}
    n(\mathbf{r}) \xrightarrow[r\rightarrow\infty]{} n_0 \, e^{-2\sqrt{-2r\varepsilon_{\text{H}}^-}}
\end{equation}
the NCAP exchange potential will tend to
\begin{equation}
    v_x^{\text{NCAP}}(\mathbf{r}) \xrightarrow[r\rightarrow\infty]{} \frac{\sqrt{2}}{3}(1-\zeta)\sqrt{-\varepsilon_{\text{H}}^-} - c/r
    \label{eq:2}
\end{equation}

The ionization potential theorem requires that the potential decays to zero, so one needs to add the constant 
\begin{equation}
v_x^{DD} = -\frac{\sqrt{2}}{3}(1-\zeta)\sqrt{-\varepsilon_{\text{H}}^-}
\label{eq:3}
\end{equation}
to the NCAP potential and the HOMO eigenvalue obtained directly from the self-consistent field KS calculation, that is 
\begin{equation}
    v_x^-(\mathbf{r}) = v_x^{\text{NCAP}}(\mathbf{r})  + v_x^{DD} \;\;\; \text{and} \;\;\; \varepsilon_{\text{H}}^- = \varepsilon_{\text{H}}^{\text{NCAP}} + v_x^{DD}
    \label{eq:4}
\end{equation}
Equation (\ref{eq:3}) can be written in terms of $\varepsilon_{\text{H}}^{\text{NCAP}}$ to obtain a quadratic equation in $v_x^{DD}$ whose solutions
\begin{equation}
    v_x^{DD\mp} = -\frac{(1-\zeta)^2}{9}\left(1 \pm \sqrt{1-18\frac{\varepsilon_{\text{H}}^{\text{NCAP}}}{(1-\zeta)^2}} \right)
    \label{eq:dd1}
\end{equation}
can be related to $\Delta_{\text{xc}}^-$ and $\Delta_{\text{xc}}^+$, and can be used to obtain the approximations
\begin{equation}
    \varepsilon_{\text{H}}^- = -I \approx \varepsilon_{\text{H}}^{NCAP} + v_x^{DD-} \;\;\; \text{and} \;\;\; \varepsilon_{\text{H}}^+ = -A \approx \varepsilon_{\text{L}}^{NCAP} + v_x^{DD+}
    \label{eq:dd2}
\end{equation}

We will use $v_x^{DD-}$ and $v_x^{DD+}$ to shift the entire occupied and unoccupied eigenvalues to yield the $G_0W_0$@NCAP-DD approach.

\subsection{$GW$ Approximation}
\begin{figure}
    \centering
    \includegraphics[width=0.35\textwidth]{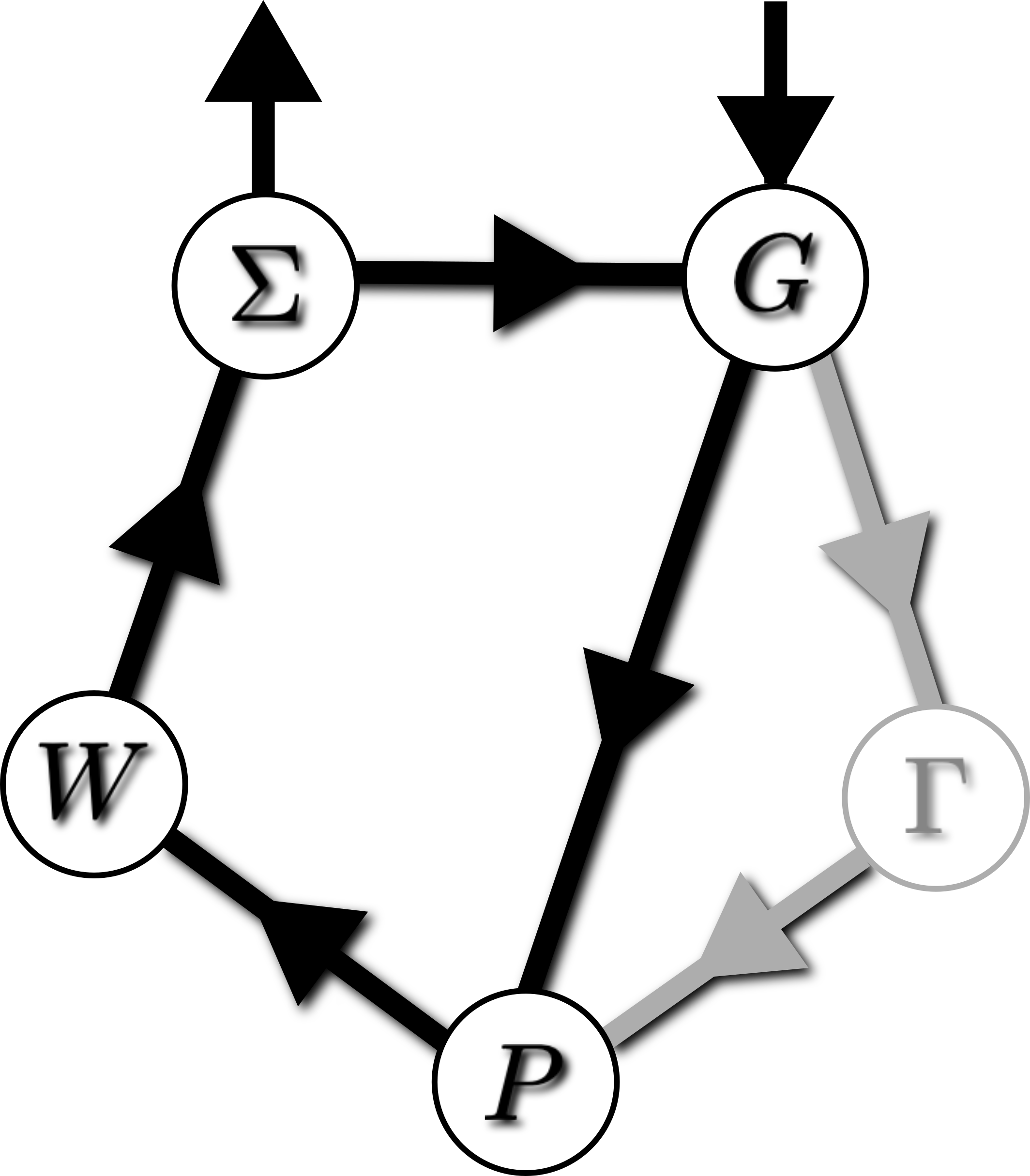}
    \caption{Schematic representation of Hedin's equations, showing the omission of vertex effects, $\Gamma$ in the $GW$ approximation. The entry point is the calculation of $G$ with some mean-field approximation. The quasiparticle energies can be obtained using the exchange-correlation self-energy effective potential. Alternatively, a better approximation to $G$ can be obtained, and the circuit iterated until self-consistency.}
    \label{fig:1}
\end{figure}

The GWA to the self-energy utilizes the one-body Green's function, $G(\mathbf{r}_1,\mathbf{r}_2,\omega)$, to describe the spectral properties of materials and molecules as its poles in the complex plane are the \emph{true} addition and removal energies of the system \cite{Hedin1965,Aryasetiawan1998,Onida2002,Martin:2016,Reining2017}. The success of the GWA depends on the ability to compute a good approximation to $G$ without an explicit need for the many-body wave function. Hedin's closed-set of equations (Figure \ref{fig:1}) 
\begin{eqnarray}
    G(1,2) &=& G_0(1,2) + \int d(1'2') \; G_0(1,1') \Sigma(1',2') G_0(2',2) \\
    P(1,2) &=& -2\iu \int d(2'2'') \; \Gamma(1; 2', 2'') G(2,2') G(2'',2) \\
    W(1,2) &=& v(1,2) + \int d(1'2') \; v(1,1') P(1',2') W(2',2) \\
    \Sigma(1,2) &=& i \int d(1'2') \; \Gamma(2';1,1') G(1',2) W(2',2) \\
    \Gamma(1; 2,3) &=& \delta(1,2)\delta(1,3) + \int d(2'2''3'3'') \; \Gamma(1; 2',3') G(2'',2') G(3',3'') \frac{\delta \Sigma(2,3)}{\delta G(2'',3'')}
\end{eqnarray}
is, in principle, an exact route that achieves precisely this goal. Different levels of approximation can be introduced into Hedin's equations, but the most natural one is neglecting all corrections to the bare vertex $\Gamma_0(1;2,3) = \delta(1,2)\delta(1,3)$ in the calculation of the polarizability $P(1,2)$ and the self-energy $\Sigma(1,2)$. Doing so yields (in simplified notation) $\Sigma = iGW$, which gives the name to this particular approximation.

Formally, Hedin's equations must be solved self-consistently. Approaches that do not do so introduce another level of approximation. In particular, the $G_0W_0$ method goes once through the loop shown in Figure \ref{fig:1}, entering with some mean-field (here assumed to be the KS-DFT solution) orbitals $\{\phi^{\text{KS}}\}$ and corresponding eigenvalues $\{\varepsilon^{\text{KS}}\}$ to compute the ``zeroth-order'' Green's function $G_0$ and screened Coulomb interaction $W_0$. The method ends once the desired quasiparticle energy $\varepsilon_p^{G_0W_0}$ is obtained by solving the non-linear quasiparticle equation given by
\begin{equation}
    \varepsilon_p^{G_0W_0} - \varepsilon_p^{\text{KS}} = \Sigma_{pp}^{\text{xc}}(\varepsilon_p^{G_0W_0}) - V_{pp}^{\text{xc}}
\end{equation}

Note that only the diagonal elements of the self-energy are important for the $G_0W_0$ method, as the orbitals are not updated. The quality of the $G_0W_0$ quasiparticle energies depends significantly on the starting KS solution due to the lack of self-consistency. A partial self-consistent approach that only enforces the self-consistency of the quasiparticle energies, known as ev$GW$, alleviates the starting-point dependence and usually improves the quality of the quasiparticle energies \cite{Shishkin2007,Stan2009}. In ev$GW$, one loops around Hedin's equations several times without updating the orbitals. As in $G_0W_0$, only the diagonal of the self-energy is needed. The updated addition and removal energies are used to compute an \emph{improved} approximation for $G$ in each loop and the method ends once the quasiparticle energies stop changing. The ev$GW$ method can be several times more expensive than the one-shot $G_0W_0$ approximation even when only one additional loop through Hedin's equations is performed. This computational burden follows from the need to solve, in principle, the quasiparticle equations for \emph{all} orbitals instead of solving them for only a handful of states \footnote{The actual relative computational cost will depend on the particularities of the $GW$ implementation, especially on how the screened Coulomb interaction $W$ is computed.}. It is therefore customary to utilize the so-called ``scissor-shift operator'' method, wherein a limited number of quasiparticle energies above and below the Fermi level are updated. In contrast, the remaining quasiparticle energies are rigidly shifted using the information of their respective space (occupied/unoccupied). Unfortunately, the validity of this shift has been recently assessed to be very limited \cite{Rasmussen2021}.

There are several ways to implement the solution of Hedin's equations, each with its own advantages and disadvantages (including numerical stability and computational cost). We have recently implemented a scalable approach \cite{Mejia2021}, based on the contour deformation (CD) approach \cite{Godby1988}, in the open-source quantum chemistry package \textsc{NWChem} \cite{nwchem1,nwchem2}. Our implementation uses the density fitting technique \cite{Whitten:1973:4496,Dunlap:1979:3396} (also known as resolution-of-the-identity technique) to avoid computing and manipulating four-center electron repulsion integrals, and the \texttt{MINRES} solver \cite{Paige:1975:617} to avoid expensive, and numerically sensitive, matrix inversions. In the molecular CD-$GW$ method, it is customary to use the Adler-Wiser representation of the irreducible polarizability \cite{Adler:1962,Wiser:1963}
\begin{equation}
    P_0(\mathbf{r},\mathbf{r}',\omega) = \sum\limits_\sigma\sum\limits_{ia} \left( \frac{\phi_{i\sigma}^{\text{KS},*}(\mathbf{r})\phi_{a\sigma}^{\text{KS}}(\mathbf{r})\phi_{a\sigma}^{\text{KS},*}(\mathbf{r}')\phi_{i\sigma}^{\text{KS}}(\mathbf{r}')}{\omega - (\varepsilon_{a\sigma}^{\text{KS}} - \varepsilon_{i\sigma}^{\text{KS}}) + \iu \eta} - \frac{\phi_{i\sigma}^{\text{KS}}(\mathbf{r})\phi_{a\sigma}^{\text{KS},*}(\mathbf{r})\phi_{a\sigma}^{\text{KS}}(\mathbf{r}')\phi_{i\sigma}^{\text{KS},*}(\mathbf{r}')}{\omega + (\varepsilon_{a\sigma}^{\text{KS}} - \varepsilon_{i\sigma}^{\text{KS}}) - \iu \eta } \right)
\end{equation}
to obtain the dynamical dielectric function
\begin{equation}
    \epsilon(\mathbf{r},\mathbf{r}',\omega) = \delta(\mathbf{r},\mathbf{r}') - \int d\mathbf{r}'' \; v(\mathbf{r},\mathbf{r}'') P_0(\mathbf{r}'',\mathbf{r}',\omega)
\end{equation}
and, from its inverse, the screened Coulomb interaction
\begin{equation}
    W_0(\mathbf{r},\mathbf{r}',\omega) = v(\mathbf{r},\mathbf{r}') + \int d\mathbf{r}'' \; \left[ \epsilon^{-1}(\mathbf{r},\mathbf{r}'',\omega) - \delta(\mathbf{r}-\mathbf{r}'') \right] v(\mathbf{r}'',\mathbf{r}')
    \label{eq:w0}
\end{equation}

In the preceding equations, $\sigma$ labels the spin, $i$ labels occupied orbitals, and $a$ labels unoccupied ones.  Note that $W_0$, Equation (\ref{eq:w0}), was conveniently split into an exchange part (first term) and a correlation part (second term). The convolution of $G_0$ with the exchange part of $W_0$ is performed analytically to yield exchange-like potential
\begin{equation}
    \Sigma_\sigma^x(\mathbf{r},\mathbf{r}') = -\sum\limits_i \phi_{i\sigma}^{\text{KS}}(\mathbf{r})\phi_{i\sigma}^{\text{KS},*}(\mathbf{r}')v(\mathbf{r},\mathbf{r}') \;\; .
\end{equation}

\begin{figure}
    \centering
    \includegraphics[width=0.5\textwidth]{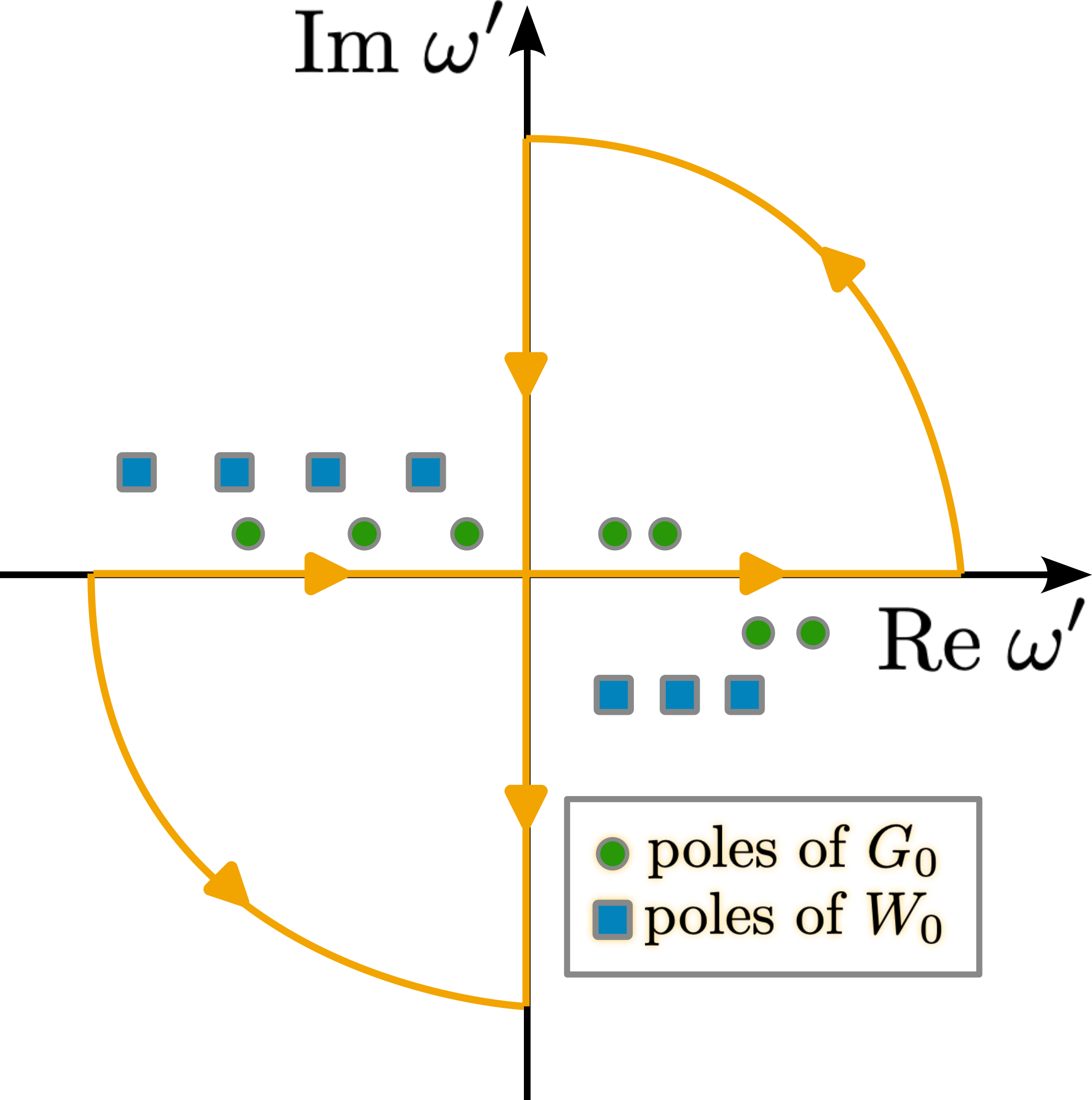}
    \caption{Schematic representation of the contour deformation technique to obtain $\Sigma_c$. The contours are chosen to contain the poles of $G_0$ but not the poles of $W_0$. The integral over the real frequencies $\omega'$ is obtained using a numerical integration over the imaginary axis and the residue theorem.}
    \label{fig:cdgw}
\end{figure}

The convolution with the correlation part is performed semi-analytically following the integration paths shown in Figure \ref{fig:cdgw}, wherein the contours are taken to infinity and chosen to exclude the poles of $W_0$. At infinitely large $\omega'$, the product $G_0(\mathbf{r},\mathbf{r}',\omega+\omega')W_0(\mathbf{r},\mathbf{r}',\omega')$ vanishes, so the real-frequency integral can be computed by subtracting the imaginary-frequency integral from the contour integrals. The integration along the imaginary axis is performed numerically, whereas the contour integrals are replaced by a sum over the residues enclosed within the contours. More details about the CD-$GW$ approach implemented in \textsc{NWChem} can be found in Reference \onlinecite{Mejia2021}.

\section{Computational Methods}
The $G_0W_0$@NCAP-DD approach was assessed using the GW100 \cite{vanSetten2015} dataset to compare vertical ionization potentials and electron affinities for a series of small molecules. All calculations were performed using the current release of \textsc{NWChem} (version 7.2.2)\footnote{\href{https://github.com/nwchemgit/nwchem/releases}{https://github.com/nwchemgit/nwchem/releases}}. 

The polarization-consistent pcSseg-3 basis set \cite{Jensen2015} was selected due to its balanced description between valence addition/removal energies and core removal energies within the GWA \cite{Mejia2022}. Since there is no optimal fitting basis to pair with pcSseg-3, we automatically generated it following the procedure detailed in Reference \onlinecite{Stoychev:2017:554}. We used the def2-QZVP/def2-universal-JKFIT basis set \cite{Weigend:2003:12753,JKFIT} combination for all those elements for which the pcSseg-3 basis set is not defined. 

For CD-$GW$, the integral over the imaginary frequencies was discretized using a Gauss-Legendre quadrature with 200 points and the imaginary infinitesimal was set to 0.001. The quasiparticle energies were obtained by solving the corresponding non-linear quasiparticle equations using Newton's fixed-point approach.
To reduce numerical noise as much as possible (especially for ev$GW$ calculations), we used the \textsc{SIMINT} electron repulsion integral package \cite{Pritchard2016} with no prescreening of the primitive products, a Schwarz-screening threshold of $10^{-16}$, tight SCF convergence of $10^{-9}$ Hartrees, and a \texttt{xfine} numerical integration grid\footnote{See \textsc{NWChem} documentation for more details about these settings.}. 

Following Section \ref{sec:theory}, the calculations labeled as NCAP-DD used a shift based on the estimated NCAP derivative discontinuity given by Equation (\ref{eq:dd1}), 
\begin{equation}
    \varepsilon_i^{\text{NCAP-DD}} = \varepsilon_i^{\text{NCAP}} + v_x^{\text{DD}-}  \;\; \text{and} \;\; \varepsilon_a^{\text{NCAP-DD}} = \varepsilon_a^{\text{NCAP}} + v_x^{\text{DD}+} \;\; ,
\end{equation}
The shift was removed while solving the quasiparticle equations, i.e.
\begin{eqnarray}
    V_{ii}^{\text{xc}} &=& \int d\mathbf{r} \; \phi_i(\mathbf{r}) v_{\text{xc}}(\mathbf{r}) \phi_i(\mathbf{r}) - v_x^{\text{DD}-} \\
    V_{aa}^{\text{xc}} &=& \int d\mathbf{r} \; \phi_a(\mathbf{r}) v_{\text{xc}}(\mathbf{r}) \phi_a(\mathbf{r}) - v_x^{\text{DD}+} \; ,  
\end{eqnarray}

Partially self-consistent ev$GW$ calculations were performed by computing the screened Coulomb interaction with the so-called spectral decomposition approach (see Reference \onlinecite{Mejia2021} for more details). This method scales as $\mathcal{O}(N^6)$ but, many times, is more stable than the CD-$GW$ method.

\section{Results and Discussion}
\subsection{Vertical Ionization Potentials}
We will start by comparing VIPs obtained with $G_0W_0$ and ev$GW$ using three different starting points: NCAP-DD, NCAPR-DD, and PBE. Figure \ref{fig:evgw} shows that all but one ev$GW$ VIPs differ less than 0.1 eV, most of which lie between $\pm 0.05$ eV (shaded area). The largest difference, 0.11 eV, is observed for the VIP of NaCl predicted by ev$GW$@NCAP-DD and ev$GW$@NCAPR-DD (-0.058 eV and +0.053 eV deviations relative to ev$GW$@PBE, respectively). The order of these deviations is in line with those reported in the literature for VIPs when comparing ``pure'' DFAs only \cite{vanSetten2013}. DFAs with some admixture of exact exchange show slightly different behaviors that can push the overall starting-point dependence to values as high as $0.5$ eV \cite{Marom2012}. 

\pgfplotstableread{gw100.dat} \homotable

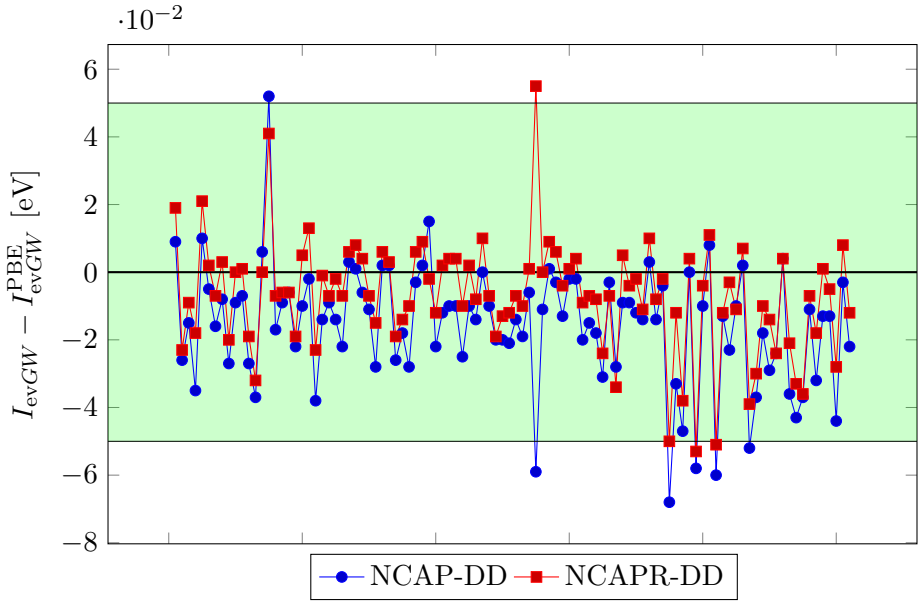
\begin{figure}
\centering
\begin{tikzpicture}
    \begin{axis}[%
    width = 0.75\textwidth,
    height = 0.5\textwidth,
    legend style = {at={(0.25,-0.02)}, anchor=north west},
    legend columns=-1,
    xticklabels = {,,},
    ylabel = {$I_{\text{ev}GW} - I_{\text{ev}GW}^{\text{PBE}}$ [eV]}
    ]
    \addplot table[y expr = \thisrow{evGW_NCAP}-\thisrow{evGW_PBE}, x = ID] from \homotable ;
    \addplot table[y expr = \thisrow{evGW_NCAPR}-\thisrow{evGW_PBE}, x = ID] from \homotable ;
    \draw[color=black, thick] (axis cs:\pgfkeysvalueof{/pgfplots/xmin},0) -- (axis cs:\pgfkeysvalueof{/pgfplots/xmax},0);
    \draw[name path=b, color=black, ultra thin] (axis cs:\pgfkeysvalueof{/pgfplots/xmin},0.05) -- (axis cs:\pgfkeysvalueof{/pgfplots/xmax},0.05);
    \draw[name path=a, color=black, ultra thin] (axis cs:\pgfkeysvalueof{/pgfplots/xmin},-0.05) -- (axis cs:\pgfkeysvalueof{/pgfplots/xmax},-0.05);
    \addplot[color=green, fill=green, fill opacity=0.2] fill between [ of=a and b] ;
    \legend{NCAP-DD, NCAPR-DD};
    \end{axis}
\end{tikzpicture}
\caption{Deviation of the self-consistent ev$GW$ vertical ionization potentials, in eV, relative to the ev$GW$@PBE value. The $x$-axis is ordered according to the list of molecules shown in the Supplementary Material.}
\label{fig:evgw}
\end{figure}

\begin{figure}
\includegraphics[width=0.75\textwidth]{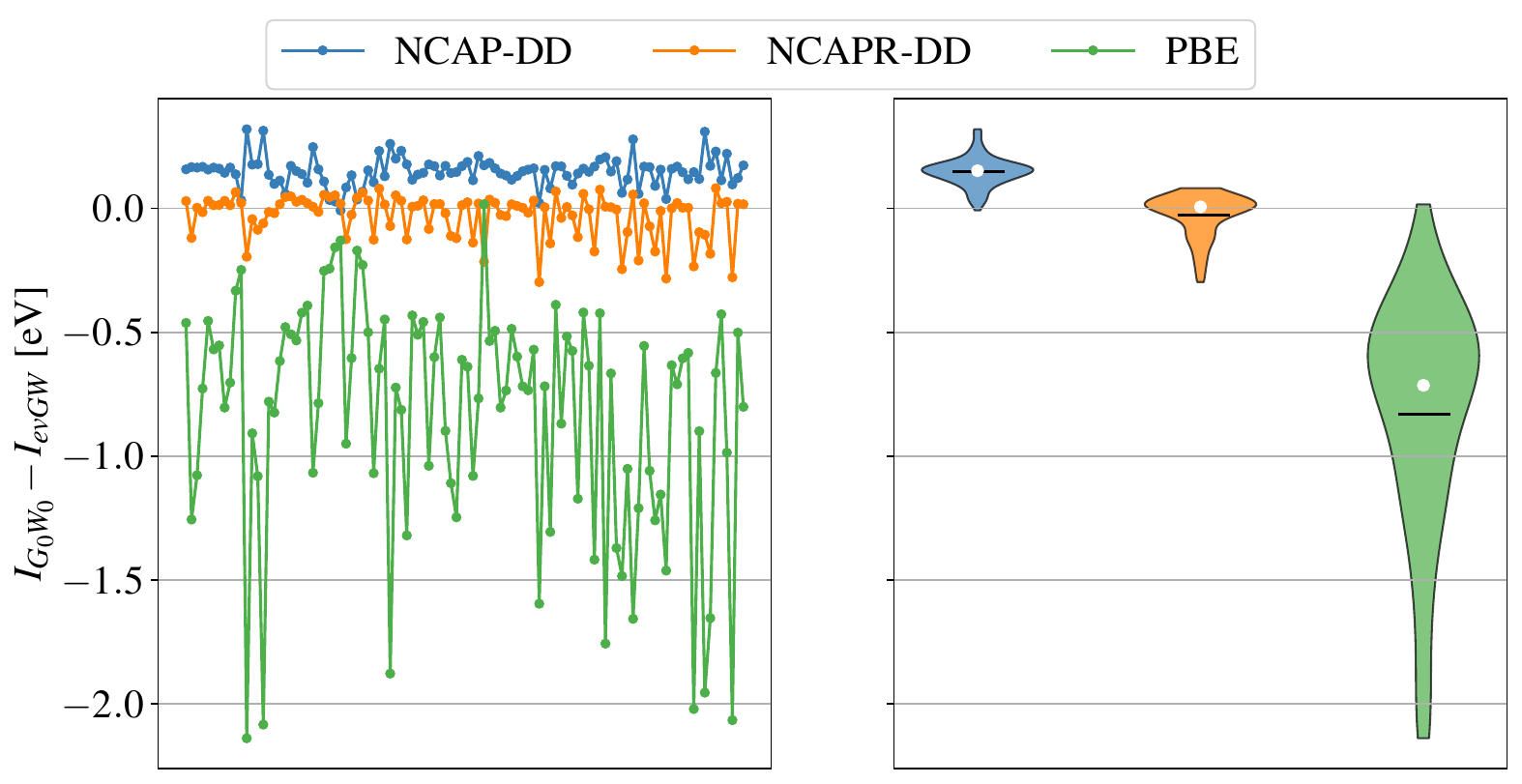}
\caption{Vertical ionization potential difference, in eV, between $G_0W_0$@DFA and the corresponding self-consistent ev$GW$@DFA one. The left plot shows individual differences with the $x$-axis ordered according to the list of molecules shown in the Supplementary Material. The violin plots on the right show the aggregated statistics with the mean appearing as a horizontal line and the median as a white dot.}
\label{fig:VIPshift}
\end{figure}

Having established the equivalency of the ev$GW$ ending points for the three starting points, we now turn to compare the VIP shift going from $G_0W_0$ to self-consistent ev$GW$ for the same starting points. Figure \ref{fig:VIPshift} shows that $G_0W_0$@PBE VIPs underestimate the ev$GW$@PBE values by as much as 2 eV. In contrast, $G_0W_0$@NCAP-DD and $G_0W_0$NCAPR-DD VIPs are much closer to the corresponding ev$GW$ values. The modest revisions introduced in NCAPR \cite{Carmona2022:1} translate into a non-negligible improvement for the $G_0W_0$ VIPs as the mean absolute deviation diminishes from 0.149 eV for $G_0W_0$@NCAP-DD to just 0.061 eV for $G_0W_0$@NCAPR-DD. It is worth mentioning, however, that the average improvement comes with a slight degradation in the dispersion of the errors, with the standard deviation increasing from 0.059 eV (NCAP-DD) to 0.088 eV (NCAPR-DD).  

Based on these results, it is possible to conclude that $G_0W_0$@NCAP-DD and, especially, $G_0W_0$@NCAPR-DD VIPs are no different than those obtained with ev$GW$ starting from other KS solutions using ``pure'' DFAs. 

\subsection{Vertical Electron Affinities}
A similar analysis can be performed for the electron affinities of the molecules contained in the GW100 dataset. Figure \ref{fig:evgwlumo} shows that the ev$GW$ starting-point dependence is stronger for VEAs than for VIPs, as some values deviate as much as 1 eV. These rather large differences might be due to the qualitatively different behavior of the exchange potentials of the NCAP DFAs, which decay as $-c/r$, from other standard GGA potentials which decay much faster. The shapes of the orbitals in the unoccupied space, including the LUMO, seem to be more affected by the nearly correct asymptotic decay of the NCAP potential. However, further explorations are needed to conclusively determine the origin of the VEA deviations. We note that a closely related DFA with a correct asymptotic potential exhibits similar effects on the unoccupied part of the spectrum, although the effect was attributed to an odd behavior in its potential at intermediate distances \cite{Carmona2015}.

Unfortunately, there is a lack of high-level reference VEAs values for all the molecules contained in the GW100 dataset. Comparison to the experiment is also not straightforward as, most of the time, the experiments measure adiabatic electron affinities (with rather large error bars). A further complication is that the majority of VEAs in the GW100 dataset are negative, meaning that the corresponding anions are unbound and, consequently, that experimental data is difficult to obtain. Because of this, we will set aside the question of which ev$GW$@DFA limit is better for VEAs.

\pgfplotstableread{gw100_lumo.dat} \lumotable

\begin{figure}
\centering
\begin{tikzpicture}
    \begin{axis}[%
    width = 0.75\textwidth,
    height = 0.5\textwidth,
    legend style = {at={(0.25,-0.02)}, anchor=north west},
    legend columns=-1,
    xticklabels = {,,},
    ylabel = {$A_{\text{ev}GW} - A_{\text{ev}GW}^{\text{PBE}}$ [eV]}
    ]
    \addplot table[y = NCAP, x = ID] from \lumotable ;
    \addplot table[y = NCAPR, x = ID] from \lumotable ;
    \draw[color=black, thick] (axis cs:\pgfkeysvalueof{/pgfplots/xmin},0) -- (axis cs:\pgfkeysvalueof{/pgfplots/xmax},0);
    \draw[name path=b, color=black, ultra thin] (axis cs:\pgfkeysvalueof{/pgfplots/xmin},0.05) -- (axis cs:\pgfkeysvalueof{/pgfplots/xmax},0.05);
    \draw[name path=a, color=black, ultra thin] (axis cs:\pgfkeysvalueof{/pgfplots/xmin},-0.05) -- (axis cs:\pgfkeysvalueof{/pgfplots/xmax},-0.05);
    \addplot[color=green, fill=green, fill opacity=0.2] fill between [ of=a and b] ;
    \legend{NCAP-DD, NCAPR-DD};
    \end{axis}
\end{tikzpicture}
\caption{Deviation of the self-consistent ev$GW$ vertical electron affinities, in eV, relative to the ev$GW$@PBE value. The $x$-axis is ordered according to the list of molecules shown in the Supplementary Material.}
\label{fig:evgwlumo}
\end{figure}
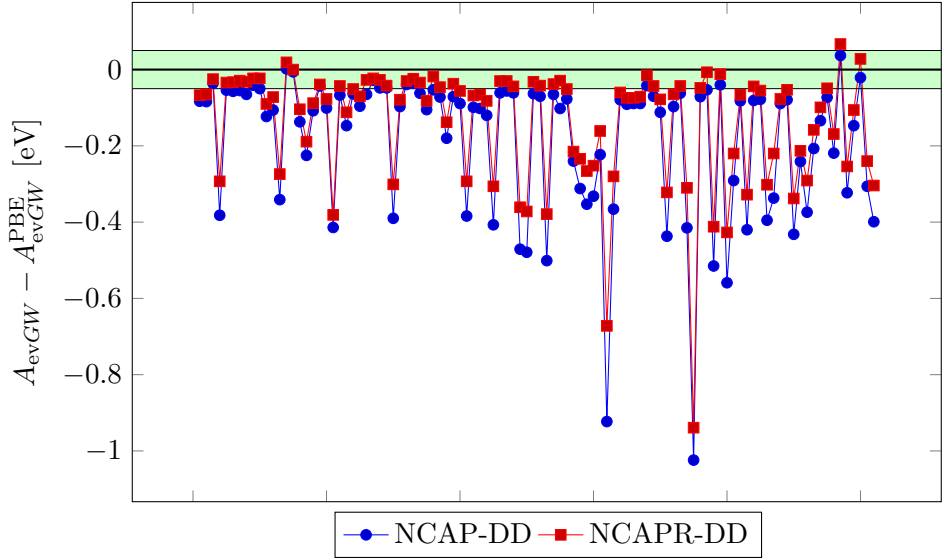

\begin{figure}
\includegraphics[width=0.75\textwidth]{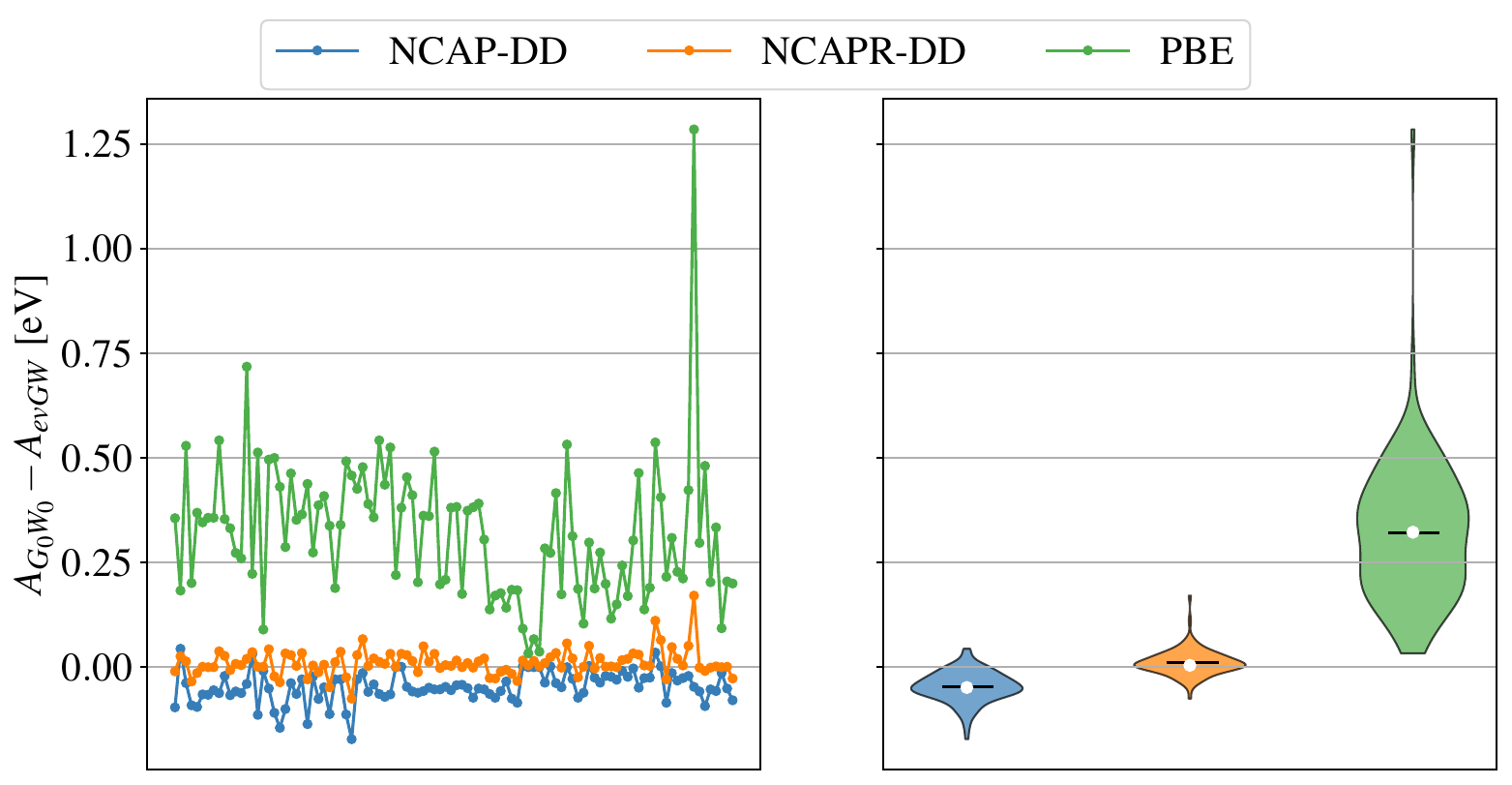}
\caption{Vertical electron affinity difference, in eV, between $G_0W_0$@DFA and the corresponding self-consistent ev$GW$@DFA one. The left plot shows individual differences with the $x$-axis ordered according to the list of molecules shown in the Supplementary Material. The violin plots on the right show the aggregated statistics with the mean appearing as a horizontal line and the median as a white dot.}
\label{fig:VEAshift}
\end{figure}

 We can, nevertheless, assess the quality of the $G_0W_0$ VEAs relative to the corresponding ev$GW$ values. Figure \ref{fig:VEAshift} shows that $G_0W_0$@NCAP and $G_0W_0$@NCAP-DD VEAs are very close to the respective ev$GW$ results. The mean absolute deviations are just 0.049 eV and 0.021 eV for $G_0W_0$@NCAP and $G_0W_0$@NCAP-DD, respectively. Different from the VIP case, $G_0W_0$@PBE VEAs are relatively close to the corresponding ev$GW$@PBE VEAs but systematically overestimated. 
 
 It is well-known that updating $W$ in (partial) self-consistent $GW$ calculations often results in underscreening since quasiparticle energy differences appear in the denominator of $W$ \cite{Caruso2016,Knight2016}. This underscreening leads to photoelectron spectra that appear elongated, with overestimated ionization potentials and underestimated electron affinities. The fact that $G_0W_0$@NCAP-DD overestimates the VIPs and underestimates the VEAs relative to ev$GW$@NCAP-DD might be problematic because this will translate into overshooting an already overestimated fundamental gap. This effect is almost absent in $G_0W_0$@NCAPR-DD.

\subsection{Comparison to other methods}

One of the main advantages of using the NCAP-DD solution as a starting point is that Equation \ref{eq:dd1} is valid for any molecular system. In contrast, the range separation and exact exchange mixing parameters must be changed from system to system when using optimally tuned range-separated hybrids (OTRSH), often needing several self-consistent calculations to achieve the desired outcome. Furthermore, the NCAP ground-state solution can be obtained very efficiently by using density fitting to approximate the Coulomb potential. This is very important because the ground-state solution is often the bottleneck of the whole $G_0W_0$@DFA approach when only a handful of frontier quasiparticle energies are of interest. As a consequence, $G_0W_0$@NCAP-DD calculations can be significantly faster than OTRSH and $G_0W_0$@OTRSH ones. We also note the existence of a one-shot tuning, for global and range-separated hybrids \cite{Jing2022}, which should be more computationally efficient than the optimal tuning in OTRSHs. This novel one-shot approach, however, still needs tuning parameters for each individual system.

The same non-universality is found for the screened Koopman's compliant (KC) DFAs, where the screening parameter should be obtained for each orbital. In the original implementation, each parameter was obtained from separate calculations on the neutral and singly-ionized systems \cite{Borghi2014}. Approximate screening parameters can also be obtained using perturbation theory \cite{Colonna2018}, reducing the number of calculations needed. 

Another related method is the $\bar{\Delta}GW$ approach from Vl\v{c}ek and cols. \cite{Vlcek2018}. However, $\bar{\Delta}GW$ needs a $G_0W_0$ calculation to estimate the size of the occupied and unoccupied shifts. In contrast, NCAP-DD provides the shifts before entering the $GW$ calculation itself. However, we note that both approaches need only one iteration through Hedin's cycle to achieve the desired outcome. Another key point is that the $\bar{\Delta}GW$ method offers ionization potentials that deviate several tenths of eV from the ev$GW_0$ target. We haven't tested our approach in the ev$GW_0$ framework, but we expect to achieve the same fidelity as for ev$GW$, namely less than 0.1 eV difference between the $G_0W_0$ and ev$GW_0$.

\begin{figure}
    \centering
    \includegraphics[width=0.75\textwidth]{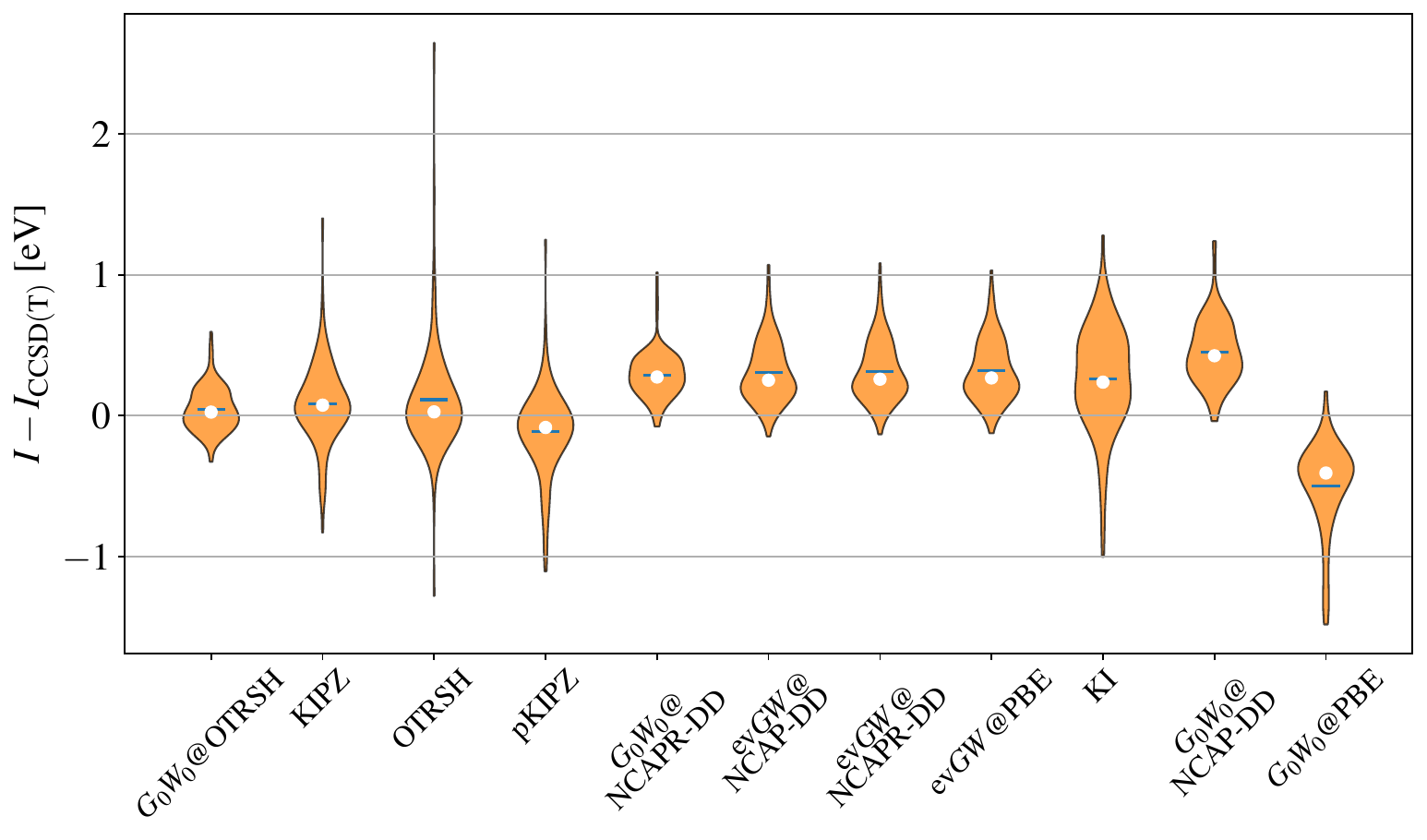}
    \caption{Comparison of the performance of several electronic structure methods for the prediction of vertical ionization potentials. The data for OTRSH and $G_0W_0$@OTRSH was obtained from Reference \cite{McKeon2022}, for KI, pKIPZ, and KIPZ Koopman's-compliant functionals from Reference \cite{Colonna2019}, and for CCSD(T) from Reference \cite{Bruneval2021}. }
    \label{fig:comparison}
\end{figure}

For completeness, Figure \ref{fig:comparison} compares the accuracy of the VIPs obtained with an OTRSH DFA \cite{McKeon2022}, several KC DFAs \cite{Colonna2019}, and several flavors of $GW$ calculations, relative to a CCSD(T) reference \cite{Bruneval2021}. The methods are ordered by increasing mean absolute error. Is not surprising that several system-dependent approaches are the best performing of the set, but the ev$GW$ approaches, including the $G_0W_0$@NCAP-DD family, perform rather well. It is unclear what performance would the eV$GW$@OTRSH, but the VIPs will likely be overestimated due to the underscreening attained while iterating $W$. 

Finally, we will only mention that the renormalized singles approach goes beyond the $G_0W_0$@NCAP-DD family (and the related $\bar{\Delta}GW$), as the underlying KS orbitals are also updated before the full $GW$ calculation.

\section{Conclusions and Future Outlook}
We have shown that an estimate of the derivative discontinuity (DD) present in the NCAP family of functionals can be used, at no computational cost, to provide a remarkably good starting point for $G_0W_0$ calculations. The DD estimate is split into two shifts, one for the occupied part of the spectrum and one for the unoccupied part of the spectrum, according to the procedure detailed by Carmona-Esp\'indola and cols. in References \cite{Carmona2020:1} and \cite{Carmona2022:1}. The shifted eigenvalues are then used to compute an estimate of the Green's function $G$ and the screened Coulomb interaction $W$ that are in agreement with those obtained in partially self-consistent ev$GW$ methods. This procedure leads to a one-shot $G_0W_0$ approach that yields vertical ionization potentials and electron affinities with ev$GW$ quality, saving valuable computational time.

The best performance overall was obtained using the $G_0W_0$NCAPR-DD approach, showing mean absolute deviations, relative to ev$GW$@NCAPR-DD, of 0.06 eV and 0.02 eV for ionization potentials and electron affinities, respectively. The $G_0W_0$@NCAP-DD performance is not far behind, although it can aggravate the fundamental gap overestimation seen in ev$GW$ approaches.

Vertical electron affinities show a strong starting-point dependence, even with the self-consistent ev$GW$ method. Explorations to obtain the root of these differences are being conducted.

The $G_0W_0$@NCAP-DD approach is naturally well-suited to provide the quasiparticle energies and screened Coulomb interaction needed to obtain neutral excitations via the Bethe-Salpeter equation formalism.

\section{Acknowledgments}
This work was supported by the Center for Scalable Predictive methods for Excitations and Correlated phenomena (SPEC) which is funded as part of the Computational Chemical Sciences (CCS) program, under FWP 70942, by the U.S. Department of Energy, Office of Science, Office of Basic Energy Sciences, Division of Chemical Sciences, Geosciences and Biosciences at Pacific Northwest National Laboratory (PNNL). PNNL is a multi-program national laboratory operated by Battelle Memorial Institute for the United States Department of Energy under DOE contract number DE-AC05-76RL01830.

\section{Supplemental Material}
Tables with names of the systems contained in the GW100 benchmark dataset as well as individual ionization potentials and electron affinities.

\bibliography{references}

\end{document}